\documentclass[twoside,twocolumn]{article}

\usepackage{blindtext} 
\usepackage{amssymb,amsmath,times,url}

\usepackage{siunitx}
\usepackage{graphicx}

\usepackage[sc]{mathpazo} 
\usepackage[T1]{fontenc} 
\usepackage{microtype} 

\usepackage[english]{babel} 

\usepackage[hmarginratio=1:1,top=32mm,columnsep=20pt]{geometry} 
\usepackage[hang, small,labelfont=bf,up,textfont=it,up]{caption} 
\usepackage{booktabs} 

\usepackage{lettrine} 

\usepackage{enumitem} 
\setlist[itemize]{noitemsep} 

\usepackage{abstract} 

\usepackage{titlesec} 
\renewcommand\thesection{\Roman{section}} 
\renewcommand\thesubsection{\roman{subsection}} 
\titleformat{\section}[block]{\large\scshape\centering}{\thesection.}{1em}{} 
\titleformat{\subsection}[block]{\large}{\thesubsection.}{1em}{} 

\usepackage{fancyhdr} 
\usepackage{titling} 

\usepackage{hyperref} 

\pretitle{\begin{center}\Huge\bfseries} 
\posttitle{\end{center}} 
\title{Federated Learning With Highly Imbalanced Audio Data} 
\author{%
\textsc{Marc C. Green and Mark D. Plumbley} \\[1ex] 
\normalsize Centre for Vision, Speech and Signal Processing, University of Surrey \\ 
\normalsize \href{mailto:marc1701@gmail.com}{marc1701@gmail.com} \\
\normalsize \href{mailto:m.plumbley@surrey.ac.uk}{m.plumbley@surrey.ac.uk} 
}
\date{} 
\renewcommand{\maketitlehookd}{%
\begin{abstract}
\noindent Federated learning (FL) is a privacy-preserving machine learning method that has been proposed to allow training of models using data from many different clients, without these clients having to transfer all their data to a central server. There has as yet been relatively little consideration of FL or other privacy-preserving methods in audio. In this paper, we investigate using FL for a sound event detection task using audio from the FSD50K dataset. Audio is split into clients based on uploader metadata. This results in highly imbalanced subsets of data between clients, noted as a key issue in FL scenarios. A series of models is trained using `high-volume' clients that contribute 100 audio clips or more, testing the effects of varying FL parameters, followed by an additional model trained using all clients with no minimum audio contribution. It is shown that FL models trained using the high-volume clients can perform similarly to a centrally-trained model, though there is much more noise in results than would typically be expected for a centrally-trained model. The FL model trained using all clients has a considerably reduced performance compared to the centrally-trained model.
\end{abstract}
}

\begin{document}

\maketitle


\section{Introduction}
\label{s:intro}

Since the establishment of the Detection and Classification of Acoustic Scenes and Events (DCASE) challenge series in 2013 \cite{Stowell2015}, machine learning for environmental audio has become a thriving area of research. Broadly split into the dual tasks of acoustic scene classification (ASC) and sound event detection (SED), with various specialist subtasks, there are applications in audio surveillance \cite{Crocco2016}, smart home systems \cite{Krstulovic2018}, and urban acoustic environment monitoring \cite{Bello2018, Bunting2009}.

Until recently, there has been little consideration by the community of the privacy implications of these applications of the technology. Deploying sound sensor networks in public or private spaces using standard centralised machine learning creates a scenario where a great deal of identifiable and potentially sensitive personal audio information is relayed to a central server, and could become accessible to unauthorised parties.

Federated Learning (FL) \cite{McMahan2017} is one method that has been proposed to address these concerns. Using FL, the raw data is never transmitted from the local client devices to the server. Instead, local data is used to train local models, with only the trained parameters from these models transmitted to the central server. The server aggregates the parameters from all the client devices into a global model, which is then redistributed to the client devices, where another round of local training takes place. This exchange continues until the model is trained. Thus, client devices can benefit from other clients' data, without that local data ever leaving each device.

One of the key challenges of this approach is that there may be significant variation in the environmental circumstances of different client devices, such that the data may be unbalanced and not independent and identially distributed (IID) across devices. That is, the distribution of data at each local device may not be representative of the global distribution, in particular some devices may have much more data than others, representing different subsets of the classes of interest. This means that gradients estimated in client models may be very biased and not representative of the gradient over the full dataset \cite{Ksieh2020}. Despite these problems, FL has been deployed as part of commercial predictive text models \cite{Hard2019}.

Recently, Johnson \emph{et al.} \cite{Johnson2021} presented a study into FL for sound event detection using synthesised soundscape data. In their study, data was artificially split across 100 clients to simulate both IID and non-IID distributions, with the amount of training data evenly split across clients (300 ten-second clips per client), using a total of 25 event classes. That study found that, for IID data, FL was able to produce models performing almost as well as centralised training, but using non-IID data greatly reduced performance \cite{Johnson2021}.

In this paper, we present a study using the FSD50K dataset \cite{Fonesca2020}, a large collection of weakly-labelled real audio data, sourced from freesound.org \cite{Freesound}, featuring 200 classes, and with metadata including the usernames of the uploaders of each sound clip. We investigate the effectiveness of FL in a challenging audio scenario where, unlike in \cite{Johnson2021}, the data has not been curated especially for an FL investigation. Instead, by utlising the uploader metadata, we simulate clients based on `real' users, with the highly imbalanced distributions one might expect from real-world users kept intact. We believe this may serve to highlight the challenges inherent in this approach and indicate potential paths for future work.

This paper is organised as follows: Section \ref{s:background} establishes the key background to this work, including details on the Federated Averaging (FedAvg) algorithm (Section \ref{ss:fedavg}) and FSD50K dataset (Section \ref{ss:dataset}). Section \ref{s:method} outlines the experimental setup used, with Section \ref{s:results} presenting results from these trials. Section \ref{s:discussion} discusses these results, and concluding thoughts and ideas for future work are presented in Section \ref{s:conclusion}.

\section{Background}
\label{s:background}

\subsection{Federated Learning}
\label{ss:fedavg}
As covered in Section \ref{s:intro}, FL requires global model held on a central server device, which is distributed to several client devices. These local model copies are trained on the clients using the data available locally. There are periodic rounds of communication between clients and server, where clients transmit the locally-calculated model parameters back to the server. The key component of FL is the incorporation by the server of these parameters from the client models back into the global model. Several approaches have been proposed to do this, including Federated Stochastic Variance Reduced Gradient (FSRVG) \cite{Konecny2016}, and CO-OP \cite{Wang2017}, but the most common is the FedAvg algorithm \cite{McMahan2017}, shown to have better performance than the others \cite{Nilsson2018}.

The FedAvg algorithm has three key parameters \cite{McMahan2017, Nilsson2018}:
\begin{itemize}
  \item $C$ - the fraction of clients selected for local training for each round of communication between central server and clients.
  \item $E$ - the number of local epochs.
  \item $B$ - the local mini-batch size.
\end{itemize}

\noindent To start, a global model $w_t$ is initialised with random parameters. A subset of clients $S_t$ is randomly selected according to $C$. At each time step $t$, the local models $w^k_t$ for client $k$ in this subset are updated to the global model, i.e. $w^k_t \leftarrow w_t$. Each client computes $E$ epochs of training over a mini-batch of size $B$ per communication round to yield new local models $w^k_{t+1}$. These local models are transmitted to the server, which aggregates them into a new global model according to \cite{Nilsson2018, Konecny2016}:

\begin{equation}
  \label{eq:fed}
  w_{t+1} = \sum_{k\in S_t} \frac{n_k}{\mu_t}w^k_{t+1}
\end{equation}

\noindent where $\mu_t$ is the total number of data points on all clients selected in this round and $n_k$ is the number of data points on client $k$. In effect, the new global model parameters are a weighted mean of local parameters from each client.

\subsection{Dataset}
\label{ss:dataset}

The dataset used in this study is the FSD50K dataset of human-labelled sound events \cite{Fonesca2020}. FSD50K features 51,197 audio clips of varying length drawn from freesound.org \cite{Freesound}, labelled with 200 classes drawn from the AudioSet ontology \cite{Gemmecke2017}. The representation of classes in the dataset is very imbalanced, with many more examples of some classes than others. There are, for instance, over 14,000 clips with a `music' label, over 5,000 labelled `human voice', but only 163 labelled `printer'. The FSD50K metadata includes the username of the freesound.org user that uploaded each sound clip, denoted `uploaders'. There are 3,647 unique uploaders in the training set, and the amount of data is highly imbalanced between these, with only a very few contributing large amounts of audio. The vast majority of uploaders contribute a small amount of audio, with 3,080 contributing 10 clips or less, 1,635 of these contributing only one clip. We have used this uploader information to divide the data into clients, retaining systematic biases including use of different recording equipment and the different classes contributed by each uploader. To keep the number of clients manageable for this initial study, we used a reduced version of the dataset limited to uploaders that contributed 100 or more clips, resulting in 57 `high-volume' clients. These 57 clients contain in total 35\% of the data included in the FSD50K training set.

\subsection{Method}
\label{s:method}
Audio from the FSD50K dataset is processed the same way as in the paper that introduced the dataset \cite{Fonesca2020}. The audio is downsampled to 22.5 kHz and transformed to a 96-band mel-spectrogram. Clips are segmented into one-second patches with 50\% overlap, and these patches are processed using 30ms frames with 10ms overlap, resulting in input arrays of shape $t\times f = 101\times96$. As in \cite{Fonesca2020}, these patches inherit clip-level labels. Class scores for each patch are therefore averaged to generate clip-level predictions. These clip-level predictions are used to calculate performance metrics.

We use the VGG-like architecture that was the best-performing model tested in \cite{Fonesca2020}. This consists of three convolutional layers of 32 filters, two of 64 filters and one of 128 filters, each layer followed by batch normalisation and ReLU activation. Groups of layers with the same number of filters are followed by $2\times 2$ max-pooling. All convolution kernels are $3\times 3$. The output of these convolutional layers is summarised by both global max pooling and global average pooling, concatenated to form input to two fully-connected layers, the first of 256 units and the second of 200, matching the size of the class vocabulary.

To generate baseline results, this model was trained in the conventional non-FL centralised manner using the Adam optimiser \cite{Kingma2017} with a learning rate of \num{5e-4}. Two versions were trained, the first using the complete FSD50K dataset, and the second using the reduced version with only the high-volume client data. Following this, a series of models were trained using FL and the high-volume client data. Local models were trained using Adam, with a local learning rate of \num{5e-4}. A grid search combinining $C$ and $E$ values was performed as follows:

\begin{itemize}
  \item $C = \{0.1, 0.3, 0.5, 0.7\}$
  \item $E = \{1, 3, 5\}$
\end{itemize}

\noindent with batch size $B$ held constant at 64. These values represent proportions from 10\% to 70\% of the clients selected in each communications round, with either 1, 3, or 5 epochs of local training. Each test was run for 50 communication rounds, with area under the precision-recall curve (PR-AUC) scores \cite{sklearn} recorded for the validation data after each round. Following this, one final model was trained using FL and the complete FSD50K dataset (using $C = 0.5$ and $E = 3$).

\section{Results}
\label{s:results}

\begin{figure}
  \centering
  \includegraphics[width=\linewidth]{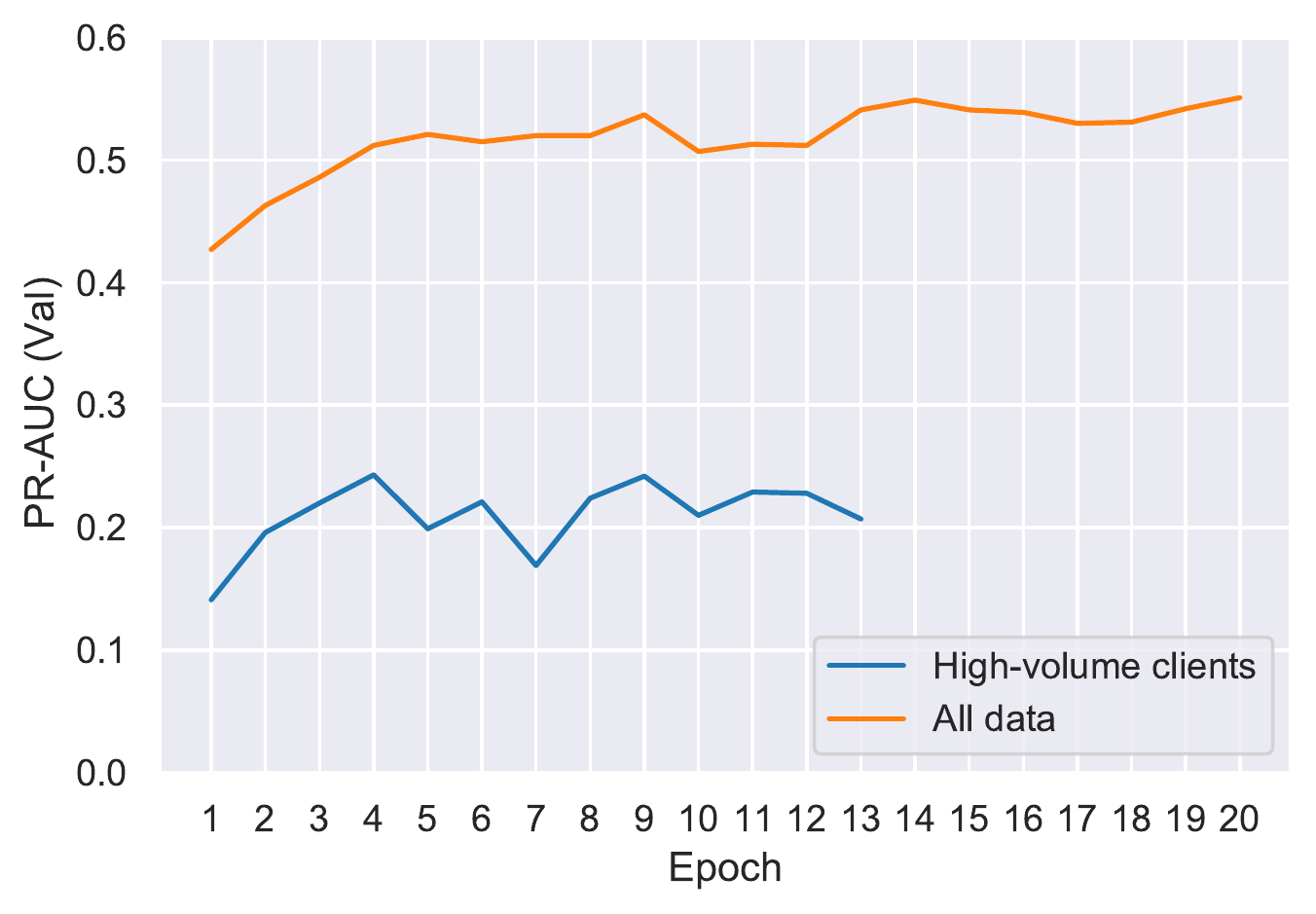}
  \caption{PR-AUC score for each epoch in the centrally-trained models.}
  \label{fi:central_model}
\end{figure}

Figure \ref{fi:central_model} shows the PR-AUC values for each epoch of training in the standard centralised models trained using both the complete dataset and the high-volume client data. Note that early stopping ended the high-volume model training after 13 epochs. It can be seen that both these models show steady improvement over the first four epochs of training before plateauing at around 0.53 and 0.23, respectively. The highest PR-AUC for the complete data is 0.55, which is similar to the benchmark centralised model results reported in \cite{Fonesca2020}. For the high-volume model, the highest PR-AUC is 0.24, which is still fairly high given that 65\% of the training data is not used in this case. These provide good baseline scores with which to compare the FL-trained models.

\begin{figure*}
  \includegraphics[width=\textwidth]{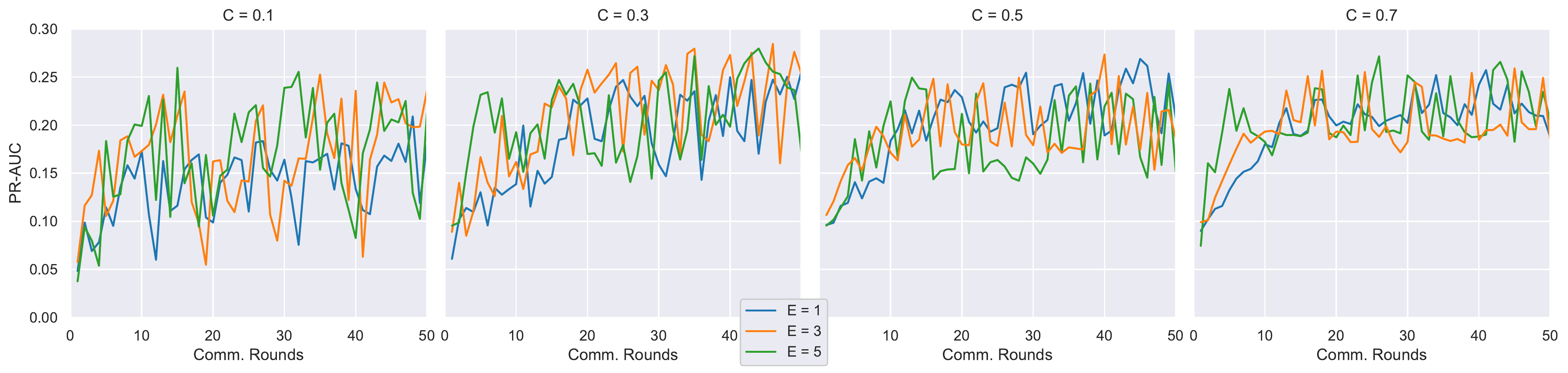}
  \caption{Global PR-AUC scores for models trained using FL across all 50 communication rounds between high-volume clients and central model.}
  \label{fi:prauc_prog}
\end{figure*}

Figure \ref{fi:prauc_prog} shows the validation PR-AUC scores achieved using FL for each central model after every round of communication with the high-volume clients. It can be seen that, though generally the PR-AUC values are much noisier than for centralised training, for all values of $C$, there are training rounds yielding PR-AUC values greater than 0.25. The very best PR-AUC score achieved in FL training is 0.284, for $C = 0.3$ and $E = 3$, though given the noisiness of the results this cannot be taken as evidence that these represent good values for those parameters. Results are less noisy when a larger fraction of clients $C$ are selected each round. For $C = 0.7$ a linear improvement in performance for the first ten communication rounds can be seen, followed by a plataeu where PR-AUC scores vary between around 0.2 and 0.25. This pattern is also somewhat evident for $C = 0.5$, though variations in performance from round-to-round are more pronounced. Results for $C = 0.3$ and $C = 0.1$ vary very erratically, though there is still an upward trend discernible for $C = 0.3$. This increased noisiness with decreasing $C$ could be due to the fact that for low values of $C$, very different subsets of data are available for training during each round of communication, thus potentially pushing the global model too far towards one subset or another in each round. When more of the data is available in each round, the bias is therefore less extreme, so the results are not as noisy. This is the main visible effect of varying $C$, and is consistent with observations in \cite{Nilsson2018}, where FL models were trained using MNIST image files \cite{Lecun1998}.

Varying the number of local training epochs $E$ seems to have only subtle effects on performance. For $C = 0.7$, the model improves more quickly in the earlier rounds for higher values of $E$, but after round 10 there is little observable effect. For $C = 0.5$, there is some indication of improved performance for $E = 3$ compared to $E = 1$ in earlier rounds, but results are too noisy to draw any conclusions. For $C = 0.1$, the maximum PR-AUC using $E = 1$ is 0.21, whereas for $E = 3$ and $E = 5$, this increases to 0.25 and 0.26, respectively. Likewise, for $C = 0.3$, maximum PR-AUC for $E = 1$ is 0.26, whereas for $E = 3$ PR-AUC is 0.27, and for $E = 5$ it is 0.28. Similar patterns can be observed when considering mean PR-AUC values. This might indicate that increasing the number of local training epochs can somewhat compensate for the smaller amounts of data used in each round for low values of $C$. For $C \geq 0.5$, however, there is no such pattern of improvement of either maximum or mean scores.

\begin{figure}
  \centering
  \includegraphics[width=\linewidth]{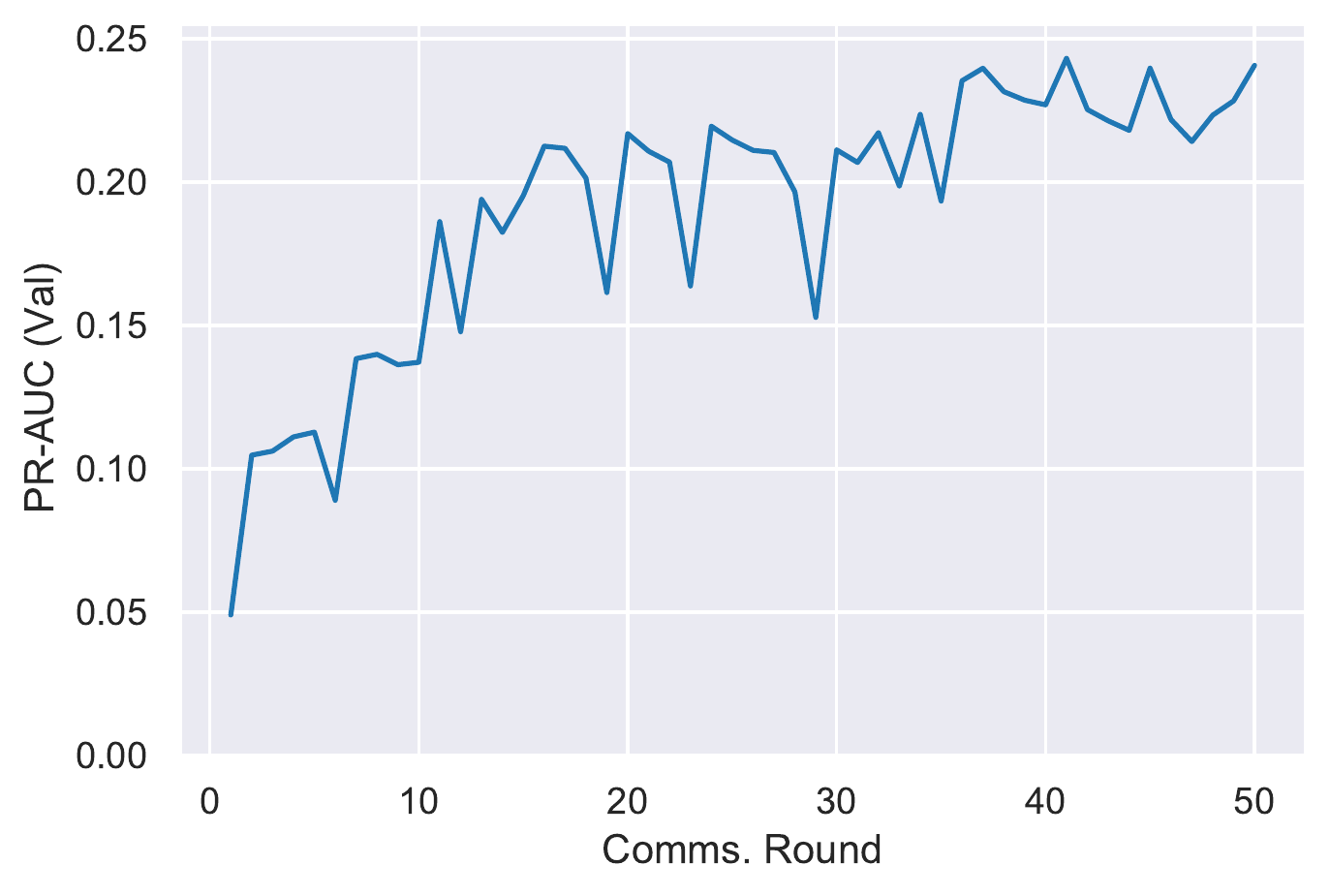}
  \caption{PR-AUC score for each communication round for a model trained using FL ($C = 0.5$, $E = 3$) and the whole FSD50K dataset.}
  \label{fi:all_data_fl}
\end{figure}

Figure \ref{fi:all_data_fl} shows the PR-AUC scores after each communication round for the FL model using clients comprising the entirety of the FSD50K dataset. As previously noted, this includes 1,635 clients with only a single point of training data each, and many more with less than 100 points of data each. PR-AUC values follow a broadly similar pattern to the high-volume $C = 0.7$ results, with a gradual improvement over early rounds, followed by a plateau where values stay around 0.2. There is, however, a small but noticeable increase at round 36 where PR-AUC scores increase from peak values of around 0.22 to 0.24.

It is very interesting that these results for an FL model trained using the whole of FSD50K are similar to the FL performance using the large uploader data only, and much lower than those achieved using all this data in a centralised manner (see Figure \ref{fi:central_model}), despite nominal access to a lot more data. Given that, upon incorporation into the global model, parameters from client models are weighted based on the amount of data in the local model (see Equation \ref{eq:fed}) it is possible that contributions from low-volume local models are too small to have an impact on global parameters when presented alongside parameters from high-volume clients.

Alternatively, since just under half of clients contribute only a single clip, and 84\% of clients contribute less than 10 clips, since clients are selected uniformly, regardless of the amount of data they contribute, it is likely that there will be considerably \emph{less} data contributing to training the global model in any given round. In fact, since the high-volume clients represent only 1.6\% of all clients, the chances of any client from this group being selected for local training over 50 communication rounds is only 40\%. It is conjectured that the small increase in PR-AUC at round 36 may have been due to one of the high-volume clients being selected in that round.

\section{Discussion}
\label{s:discussion}

Taken together, the results for the high-volume client data and the whole dataset suggest that when there are very considerable imbalances in the amount of data contributed by each client, it may take many more rounds of communication to approach the performance of a centrally-trained model than when all clients contribute substantial amounts of data, as simulated in the high-volume client models. For instance, although noisy, especially for lower values of $C$, within around 10 rounds of communication, models trained using FL and the high-volume client data are able to match the performance of the model trained centrally using this data. By contrast, the FL model trained using the complete dataset has reached only about half of the performance of the centralised model after 50 rounds of communication. It is possible that, given many more communication rounds, the performance of this model might approach that of the centralised model, but given in most practical FL scenarios, communication represents the largest cost and potential bottleneck \cite{Li2020}, this scenario is not ideal.

One way to improve performance in this case might be to implement a means by which some high-volume clients could be guaranteed to be selected in each training round, making sure a majority of the available data could be seen in a smaller number of communication rounds. Such a system, may, however, reduce the privacy of clients contributing a lot of data, as some form of additional record would be needed regarding which clients these were.

The noisiness of the results from round to round, especially for low values of $C$ in the high-volume client models, incidicate that for stable performance it might be advisable to keep the proportion of clients participating in each round high, if possible. If speed of model training is not a priority, a system could perhaps be configured to wait until a majority of clients have an update ready before performing global averaging. Unlike $C$, $E$ can be increased without increasing communication costs. The results presented here show that for lower values of $C$, increasing $E$ can result in slightly higher maximum scores achieved, whilst for higher values of $C$ this results in faster improvements in early communications rounds. In neither case does altering $E$ seem to have any effect on the noisiness of the results.

One way to reduce the varibility in results from round to round with a low $C$ might be to introduce some form of smoothing at the global level by reducing the impact of the client parameters in any particular communication round. In \cite{Nilsson2018}, an approach is tested whereby the central server retains copies of the most recent parameters from all clients, not just those participating in the current round. The global model is updated based on all of these parameters. This approach is shown to produce less noisy results between rounds, but also takes longer to converge than the updates described in Equation \ref{eq:fed}. There is also the potential of reduced privacy, in that parameters related to particular clients are retained on the server, rather than being removed immediately following calculation of the new global model, which could conceivably allow, for instance, inference attacks \cite{Jia2019}.

Another approach that could be tested would be the selection of clients in proportion to the amount of data they contribute, rather than uniformly at random. In the case of this dataset, however, this could result in the many small-volume clients hardly ever being selected, which would result in essentially replicating training using high-volume clients only. Some form of hybrid selection process might be required to balance these factors.

\section{Conclusions and Future Work}
\label{s:conclusion}

In this paper, we have presented a study in federated learning with highly imbalanced audio data derived from the FSD50K dataset. The dataset was split into clients based on upload contributors. Our results show that when all client devices contain substantial amounts of data, training using federated learning can yield global models with performance comparable to centrally-trained models, without the global model requiring access to any local data. On the other hand, when a majority of clients contribute very small amounts of data, training using federated learning can result in a performance reduction relative to a centrally-trained model, at least using the number of communications rounds tested here. It is shown that the performance of a global FL model tends to be more variable between rounds of training than would typically be expected from a centrally-trained model. This is especially true when training using small proportions of clients per round.

A possible next step would be to focus on a close investigation of the causes of the reduced performance when using the complete FSD50K development set to train a global model using federated learning, relative to training centrally. If performance can indeed be improved by making sure some high-volume clients are included in each round, the privacy implications of this should be explored in depth.

It might also be interesting to investigate the benefits of FL at the client level, rather than assessing performance using a global validation set. As mentioned in \cite{Wang2019}, a local model too dependent on local data can degrade performance by overfitting. Personalisation of the global model to local circumstances is an approach that might maximise performance for individual users. To investigate this using the full FSD50K data would necessitate amalgamation of some of the small-volume uploaders into larger clients as it is not possible to meaningfully assess performance on a single data point.

\section{Acknowledgements}
The authors would like to thank Andres Fernandez for his techincal assistance. This work was supported by grant EP/T019751/1 from the Engineering and Physical Sciences Research Council (EPSRC).

\section{Supplementary Materials}
The full Python code used to generate these results can be found at: \url{https://github.com/marc1701/fsd_fed}.

\bibliographystyle{siam}

\bibliography{mybib}


\end{document}